\documentstyle[12pt]{article}

\def\be{\begin{equation}}
\def\ee{\end{equation}}
\def\bea{\begin{eqnarray}}
\def\eea{\end{eqnarray}}
\def\ov{\overline}
\def\nab{\bigtriangledown}

\def\nnu{\nonumber}
\def\la{\label}
\def\br{\nonumber \\}
\def\h{\hat}

\begin{document}
\newcommand{\Psl}{\not\!\! P}
\newcommand{\dsl}{\not\! \partial}
\newcommand{\half}{{\textstyle\frac{1}{2}}}
\newcommand{\for}{{\textstyle\frac{1}{4}}}
\newcommand{\eqn}[1]{(\ref{#1})}
\newcommand{\npb}[3]{ { Nucl. Phys. B}{#1} ({#2}) {#3}}
\newcommand{\pr}[3]{ { Phys. Rep. }{#1} ({#2}) {#3}}
\newcommand{\plb}[3]{ { Phys. Lett. B}{#1} ({#2}) {#3}}
\newcommand{\prl}[3]{ { Phys. Rev. Lett. }{#1} ({#2}) {#3}}
\newcommand{\prd}[3]{ { Phys. Rev. D}{#1} ({#2}) {#3}}
\newcommand{\mpl}[3]{ { Mod. Phys. Lett.} {#1} ({#2}) {#3}}
\newcommand{\hepth}[1]{ [{\bf hep-th}/{#1}]}
\newcommand{\grqc}[1]{ [{\bf gr-qc}/{#1}]}

\def\lla{\mathrel{\mathpalette\fun <}}
\def\a{\alpha}
\def\b{\beta}
\def\g{\gamma}\def\G{\Gamma}
\def\d{\delta}\def\D{\Delta}
\def\e{\epsilon}
\def\et{\eta}
\def\z{\zeta}
\def\t{\theta}\def\T{\Theta}
\def\l{\lambda}\def\L{\Lambda}
\def\m{\mu}
\def\f{\phi}\def\F{\Phi}
\def\n{\nu}
\def\p{\psi}\def\P{\Psi}
\def\r{\rho}
\def\s{\sigma}\def\S{\Sigma}
\def\ta{\tau}
\def\x{\chi}
\def\o{\omega}\def\O{\Omega}
\def\lagr{{\cal L}}
\def\cd{{\cal D}}
\def\k{\kappa}
\def\be{\begin{equation}}
\def\ee{\end{equation}}
\def\tz{\tilde z}
\def\tF{\tilde F}
\def\ri {\rightarrow}
\def\cf{{\cal F}}
\def\pa {\partial}
\def\bpa{\overline{\pa}}
\def\ov{\over}
\begin{flushright}
IP/BBSR/97-22\\
CPHT-S503.0497\\
hep-th/9705058\\
\end{flushright}
\bigskip\bigskip
\begin{center}
{\large\bf On the Compactification of   type IIA  String Theory}

\vskip .9 cm
{\bf 
Jnanadeva  Maharana$^{a,b}
$\footnote{ Jawaharlal Nehru Fellow\\
e-mail: maharana@cpht.polytechnique.fr, \
maharana@iopb.ernet.in} 
and Harvendra Singh$^b$
\footnote{  e-mail: hsingh@iopb.ernet.in, hsingh@iop.ren.nic.in} } 
 \vskip 0.05cm
$^a$ {\it Centre de Physique Theorique\\
Ecole Polytechnique\\
F 91128 Palaiseau } \\ 
$^b$ {\it Institute of Physics, \\
Bhubaneswar-751 005, India}
\end{center}
\bigskip
\centerline{\bf ABSTRACT}
\bigskip
\begin{quote}
The ten dimensional  type IIA  string effective action with cosmological
constant term is  
dimensionally reduced on a d-dimensional torus to derive  
lower dimensional effective action. The symmetries of the reduced effective 
 action are examined. It is shown that the resulting six dimensional theory
does not remain invariant under $SO(4,4)$ symmetry whereas the reduced
action, in the absence of the cosmological constant 
 respects the symmetry as was shown by Sen and Vafa.  New class of black 
hole solutions are obtained in five and four dimensions in the presence
of cosmological constant. For the six dimensional theory, a four brane solution
is presented.
\end{quote}

\newpage

Recently, considerable progress has been made in our understanding of the
nonperturbative  features of superstring theories [1-3]. It is now
realised that the five consistent superstring theories might be 
envisioned as various
phases of a single unique theory [4]. Dualities play a cardinal role in 
revealing  the intimate connections between different
string theories in diverse spacetime dimensions and provide deeper insight
into string theory dynamics. 
We recall that the
predictions of T-duality are subject to tests in the perturbative frame work 
; whereas, the predictions and tests of S-duality are beyond the realms of
perturbation theory [5]. The p-branes, which appear as classical 
solutions of the
string effective action, have been instrumental in our understanding of
various duality conjectures in string theory [6]. The RR p-branes are 
interpreted
as D-p-branes of type II theories [7]. The type IIA string admits even D-branes, 
$p=0,2,4,6$ and type IIB theory, on the other hand, has the odd ones, i.e.
$p= 1,3,5$ with the identification that $-1$-brane is the instanton of the 
theory. Furthermore, for 10-dimensional spacetime, dual of a p-brane is the
$(6-p)$ brane and consequently, those p-branes with $ p \le 6 $, 
have duals with $ p \ge 0$. Thus, for $D=10$, the 8-brane
and 7-brane appearing in type IIA and type IIB string theories respectively
have special roles different from the other branes alluded to above.

A p-brane couples to $(p+1)$-form potential; therefore, the 8-brane will couple
to the potential $A_9$ whose corresponding field strength is the  ten form 
$F_{10}$.
In standard type IIA supergravity, the presence of the potential $A_9$ is
rather obscure. From the perspective of type IIA string theory, we know that
the theory admits 8-D-brane [7,8]. Notice that the equations of motion 
arising  
from the kinetic energy term $F^2_{10}$ only give rise to a conservation law 
and the presence of this term does not introduce any new dynamical degree of 
freedom. However, the effect of this additional term amounts to introduction of
cosmological constant, when we introduce the Poicare dual of ten form field strength instead. 
In this context it is worthwhile to mention that it had
been realised several years ago that the introduction of a four-form field
strength in four spacetime dimensions amounts to having a cosmological constant
term in that supergravity theory [9]. Romans [10], 
subsequently, constructed 
the massive ten dimensional type IIA supergravity theory and a complete 
construction was given in ref.11.

The study of type IIA superstring effective action in the presence of $F_{10}$,
or alternatively the theory with cosmological constant has  
drawn attention of
several authors [12-15]  in the recent past and 
it has been argued that the cosmological
constant takes only quantized values. We mention in passing, another 
interesting feature of the presence of cosmological constant in the four 
dimensional heterotic string effective action. It was shown that the equations
of motion are not invariant under S-duality transformations in the presence of
cosmological constant [16], whereas the equations of motion do respect the
symmetry when the constant is set to zero. Then, a weaker form of the 
naturalness criterion, due to 't Hooft [17], was invoked to argue that the
cosmological constant should remain small since when it required to vanish
there is enhancement of symmetry at the level of equations of motion,
derived from string effective action. We recall that the usual Einstein-Hilbert
action does not have any enhanced symmetry in the absence of the cosmological
constant as was recognised by 't Hooft, when he introduced the idea of
naturalness [17]. Since we expect string theory to provide answers to deep
questions in quantum gravity, it is hoped that  the
cosmological constant problem will be solved by string theory. Recently, Witten
has proposed a resolution of the cosmological constant problem [18]. The 
starting point is to envisage three dimensional theory with a string vacuum,
with unbroken supersymmetry and dilaton whose exponential is related to the
string coupling constant, $g_s$. In the weak coupling regime, the string 
perturbation theory is valid and cosmological constant vanishes due to 
unbroken supersymmetry. When one passes to the strong coupling limit,
$ g_s \rightarrow \infty$, the resulting theory is a Poincare invariant theory
in $3+1$ dimensions. The cosmological constant remains zero in this four 
dimensional theory since it continues to take vanishing value for all $g_s$.
We speculate that the stringy symmetries might provide a clue for the
resolution of the cosmological constant problem (see discussions below).  

It is well known, for the massless theory,
that type IIA compactified on $S^1$  with radius R is T-dual to type IIB
compactified on another circle with reciprocal radius [19]. Thus the issue of
compactification of massive type IIA theory to $D=9$ has been addressed in
the context of its T-duality to type IIB theory. It has been 
argued that
the ten dimensional type IIB theory, when compactified according to the
generalised Scherk-Schwarz [20] prescription, yield a massive theory in 9 
dimensions
and then one can explore the T-duality. Moreover, there have been attempts
to obtain various brane solutions in type IIA, IIB and M-theory [21-24]
and relate these
solutions in lower dimensions by adopting sequential steps in dimensional
reductions.

The purpose of this article is to present dimensionally reduced string
effective action for massive type IIA superstring action when we compactify it
on a d-dimensional torus. 
We investigate the symmetry properties of the reduced effective action.
In particular, we show that the six dimensional effective action for the
case of the massive theory does not respect the $SO(4,4)$ symmetry of
the corresponding six dimensional massless theory. Furthermore, we
 find new black hole solutions in five and
four space time dimensions from the reduced effective action in the presence
of cosmological constant term, and 
 we also present four-brane solutions in six 
dimensions.

The bosonic part of massive type IIA supergravity action,  in ten dimensions, 
is of interest to us.
 The action was introduced by Romans [10] and we write
an action in the string frame metric 

\bea
S_m= \int d^{10}x \sqrt{-g}&& \bigg[e^{-2\F} \left( R_g + 4 \ \pa_\mu\F 
\pa^\mu\F -{1\over 2\cdot3!} H_{\mu\nu\l} H^{\mu\nu\l}  \right)\br
&&-{1\over 2\cdot 2!}F_{\m\n}F^{\m\n}-{1\over
2\cdot4!}G_{\m\n\l\r}G^{\m\n\l\r} -{1\over2} m^2 \bigg],
\label{11}
\eea
where $\F$ is the dilaton field, $g_{\mu\nu}$ is the string 
$\sigma-$model metric and $m$ is the mass parameter. This action can be 
identified as the low energy limit of the type IIA string theory with $m^2$
playing the role of cosmological constant. 
 The NS-NS and R-R field strengths are defined as follows: 
\bea
&&F_{\m\n}=\pa_{[\m} A_{\n]} + m f_{\m\n}, \nnu\\
&&H_{\mu\nu\l} = \pa_\mu B_{\nu\l} + {\rm cyclic \
permutations}, \nnu\\
&& G_{\m\n\l\r}= \pa_{[\m} C_{\n\l\r]} + 2 A_{[\m} H_{\n\l\r]} +
2 m \ g_{\m\n\l\r}, 
\label{12}
\eea
where coefficients of the mass parameter  terms are $f_{\m\n}= B_{\m\n}$ and
$g_{\m\n\l\r}=B_{[\m\n} B_{\l\r]}$. 
The notation $[ \m  \n\cdots]$ 
implies the antisymmetrization of the indices. 
Note that the field strengths have mass dependent terms and are
the generalisations of their massless counterparts. The
advantage of writing massive type IIA action as in \eqn{11}  
is that the action for the massless theory can be  obtained by  taking the
limit $m\to 0$. 
 The action has
the invariance under massive `St\"uckelberg' gauge transformations
\bea
&& \d A_\m = - m \L_\m \nnu \\
&& \d B_{\m\n} = \pa_\m \L_\n -\pa_\n \L_\m \nnu\\
&& \d C_{\m\n\l}= -2m ( \L_\m B_{\n\l} + {\rm cyclic\ perms.}
).
\label{13}\eea
The above action has $N=2$
supersymmetry even though it involves mass terms.
The constant mass term
in the R-R sector of the theory  which has the interpretation of the
cosmological constant  can also be envisaged   as the dual of  10-form
filed strength alluded to earlier.  
Therefore, in ten dimensions, the appearance of $m^2$ terms provides
a clue for the presence of an 8-brane in type IIA theory with the hindsight.

Let us consider compactification of the ten dimensional effective action,
in presence of the cosmological constant term, on a $d$-dimensional torus. We
adopt the prescription of  Schwarz  and one of the authors (JM)[25]. 
The coordinates of D-dimensional
spacetime are denoted by $x^{\mu}$, whereas the rest which make the internal 
dimensions, the d-dimensional torus, are denoted as $x^\a$.
 In our notational conventions, we denote  ten-dimensional fields with hats 
over the fields as well as over the tensor indices 
$( {\hat \F},\ {\h g}_{{\h\m}{\h\n}},\ etc.)$,
while reserve the quantities without hats for D-dimensional ones. 
Furthermore, we assume that  the fields  are  independent of the 
``internal'' coordinates, $x^\a$. The  ten dimensional 
vielbein can be expressed in the  following form
${\h e}^{\h r}_{\h \m}=
\left(\begin{array}{cc} e^r_\m & A^{(1) \b}_\m E^a_\b \\0 & E^a_\a
\end{array}\right)$
and  ``spacetime'' metric $g_{\m\n}=e^r_\m \eta_{r s} e^s_\n$,
``internal'' metric $G_{\a\b}= E^a_\a \d_{a b} E^b_\b$. 
 
Thus, the ten dimensional metric components will be expressed in terms of the 
D-dimensional metric, gauge fields and scalars.  
 \be
  {\hat g}_{\m\n}=g_{\m\n}+ A^{(1) \a}_\m A^{(1)\b}_\n G_{\a\b}, \ \ \
A_{\m\a}^{(1) } =  {\hat g}_{\m\a}, \ \ \  
G_{\a\b}={\hat g}_{\a\b}, 
\label{13a}\ee
Similarly for the antisymmetric tensor field, coming from the NS-NS sector,
the decompositions are   
  \bea
&& A_{\m \a}^{(2)} = {\hat B}_{\m \a} - A_{\m}^{(1) \b} b_{\a \b},\ \ \
b_{\a\b}={\hat B}_{\a\b}, \nnu\\
&& B_{\m \n}^{(1)} = {\hat B}_{\m \n} -  A_{[\m}^{(1) \a} A_{\n] \a}^{(2) }
- A^{(1)\a}_\m A^{(1)\b}_\n b_{\a\b},
\eea
and the R-R fields can be decomposed as follows: 
\bea
&& c_{\a\b\g}= {\hat C}_{\a\b\g},\ \ \ a_\a= {\hat A}_\a \br
&& A_{\m\a\b}^{(3)}= {\hat C}_{\m\a\b} - A_{\m}^{(1) \d}c_{\a\b\d},\br
&& B^{(2)}_{\m\n\a}= {\hat C}_{\m\n\a} + A_{[\m}^{(1) \d} {\hat C}_{\n] \d\a} +
A_{\m}^{(1) \d_1}A_{\n}^{(1) \d_2} c_{\d_1 \d_2 \a}\br
&& C_{\m\n\l} = {\hat C}_{\m\n\l} -( A_{\m}^{(1) \d} {\hat C}_{\d\n\l} + 
{\rm cyclic\ perms.\ in} \ \m,\n,\l),\br
&&\ \ \ \ \ \ \ \ + ( A_{\m}^{(1) \d_1}A_{\n}^{(1) \d_2}{\hat C}_{\d_1\d_2
\l}+  {\rm cyclic\ perms \ in}\ \m,\n,\l ),\br
&&\ \ \ \ \ \ \ \ - A_{\m}^{(1) \d_1}A_{\n}^{(1) \d_2}A_{\l}^{(1) \d_3}
c_{\d_1\d_2\d_3}\br
&& A_{\m\a}^{(4)}= {\hat A}_\m -A_{\m}^{(1) \d}a_\d.
\label{14}\eea
Recall that the scalars are 
constructed in the ten dimensional theory by contracting the hat indices of
various tensors. In order to obtain tensors with unhatted indices, i.e. 
tensors in D-dimensions  we  
adopt the following prescription:
\be
{\cal H}_{\m\n..\a\b..}={\cal O}^{\h\m}_\m {\cal O}^{\h\n}_\n 
\cdots {\h {\cal H}}_{{\h\m}{\h\n}..\a\b..} 
\la{14a}\ee
where,  ${\cal O}^{\h\m}_\m= e^r_\m ~ {\h e}^{\h\m}_
r$ and ${\hat {\cal H}}_{{\h\m}{\h\n}..\a\b..}$ is a tensor in ten dimensions. 
Thus scalars constructed out of contraction of ten dimensional indices, as
is the case with kinetic energy terms in the action, can be expressed in the
following form in terms of scalars constructed out of various tensors in
D-dimensions, obtained through the dimensional reduction procedure,
\be
{\h {\cal H}}_{{\h\m}{\h\n}..}{\h{\cal H}}^{{\h\m}{\h\n}..}= 
{\cal H}_{\m\n..}{\cal H}^{\m\n..}+ n 
{\cal H}_{\m\n..\a}{\cal H}^{\m\n..\a}+ {n(n-1)\ov 
2!}{\cal H}_{\m\n..\a\b}{\cal H}^{\m\n..\a\b}  +\cdots+ 
{\cal H}_{\a\b..}{\cal H}^{\a\b..}, \la{14b}\ee
for the $ n-$form field strength. Following \eqn{14a} NS-NS field strengths 
are obtained as below, \bea
&& H^{(1)}_{\m\n\r}= \pa_{[\m} B_{\n\r]} - F^{(1)\d}_{[\m\n}A^{(2)}_{\r]\d},\br
&&H_{\m\n\a}= F^{(2)}_{\m\n\a}- 
 F^{(1)\d}_{\m\n} b_{\a\d} \br
&& H_{\m\a\b}=\pa_\m b_{\a\b}.
\la{15a}\eea
where $ F^{(i)}_{\m\n}= \pa_{[\m} \ A^{(i)}_{\n]}$.
The  Chern-Simon $(CS)$  term in eq.\eqn{12}, $ {\hat A} 
\wedge {\hat H}$, will give
\bea
&& [CS]_{\m\a\b\g}= -\left(a_\a\pa_\m b_{\b\g} + {\rm cyclic\ perms.\ of
\ \a,\b,\g}\right),\br 
&& [CS]_{\m\n\a\b}=  A_{[\m}^{(4)}\pa_{\n]} b_{\a\b} +
\left\{ a_\a ( F^{(2)}_{\m\n\b}-b_{\b\d}F^{(1)\d}_{\m\n})-\left(
\a\leftrightarrow \b\right)\right\}\br
&& [CS]_{\m\n\r\a}= A^{(4)}_{[\m} \left( F^{(2)}_{\n\r]\a}- 
 F^{(1)\d}_{\n\r]} b_{\a\d} \right)- a_\a H^{(1)}_{\m\n\r}\br 
&& [CS]_{\m\n\r\s}= A^{(4)}_{[\m } H^{(1)}_{\n\r\s]}.
\label{15}\eea
Then RR field strengths reduce as given below,
\bea
&& G_{\a\b\g\d}= 2 m\  b_{[\a\b}b_{\g\d]}\br
&& G_{\m\a\b\g}= \pa_\m c_{\a\b\g} + 2 [CS]_{\m\a\b\g}+ 2 m\ A^{(2)}_{\m[\a}
b_{\b\g]}\br 
&& G_{\m\n\a\b}= \pa_{[\m}A^{(3)}_{\n]\a\b} + F^{(1)\d}_{\m\n} c_{\d\a\b}+ 2 [CS]_{\m\n\a\b}+
2 m\ \left( B^{(1)}_{\m\n} b_{\a\b} -( A^{(2)}_{\m\a} A^{(2)}_{\n\b}
-\{\a\leftrightarrow \b)\}\right)\br
&&G_{\m\n\r\a}= \pa_{[\m}B^{(2)}_{\n\r]\a} -F^{(1)\d}_{[\m\n}
A^{(3)}_{\r]\d\a} + 2 [CS]_{\m\n\r\a}+2 m B^{(1)}_{[\m\n}A^{(2)}_{\r]\a}\br
&&G_{\m\n\r\s}=  \pa_{[\m }C_{\n\r\s]} +F^{(1)\d}_{[\m\n}
B^{(2)}_{\r\s]\d}+ 2 [CS]_{\m\n\r\s}+ 2 m  B^{(1)}_{[\m\n} B^{(1)}_{\r\s]}\br
&& F_{\a\b}=m b_{\a\b}\br
&& F_{\m\a}=\pa_\m a_\a + m A^{(2)}_{\m\a}\br
&& F_{\m\n}= F^{(4)}_{\m\n} + F^{(1)\d}_{\m\n} a_\d +m B^{(1)}_{\m\n}.
\label{16}\eea

Now, to consider a specific case, let us look at  the reduced effective action
in six spacetime dimensions. We utilise the identity  
 \eqn{14b} and use various definitions from 
eqs.\eqn{15a} and \eqn{16} to  write down the six-dimensional massive IIA 
action, 
\bea \int d^6x \sqrt{-g}&& \bigg[ e^{-2\f}[\ R + 4\pa_\m\f \pa^\m\f
-{1\over 12} H^{(1)}_{\m\n\l}H^{(1)\m\n\l} +{1\over 8}
Tr\pa_\m M^{-1}\pa^\m M -{1\ov 4}F^{(i)}_{\m\n} M^{-1}_{i j}
F^{(j)\m\n}\ ]\br
&& -\sqrt{G}{\Large\{ }    {1\over 2\cdot 2!}\left( ( F^{(4)}_{\m\n} +
      F^{(1)\d}_{\m\n} a_\d + m B^{(1)}_{\m\n} )^2 + 2 (\pa_\m a_\a + m
      A^{(2)}_{\m\a} )^2 + ( m \ b_{\a\b} )^2 \right) \br
&& \hskip .5in +{1 \over 2 \cdot 4!} [ 
     ( \pa_{[\m }C_{\n\r\s]} +F^{(1)\d}_{[\m\n}
     B^{(2)}_{\r\s]\d} + 2 A^{(4)}_{[\m } 
H^{(1)}_{\n\r\s]}+ 2 m B^{(1)}_{[\m\n} B^{(1)}_{\r\s]} )^2 \br
  && \hskip .5in   + 4 ( \pa_{[\m}B^{(2)}_{\n\r]\a} -F^{(1)\d}_{[\m\n}
     A^{(3)}_{\r]\d\a} + 2 A^{(4)}_{[\m} \left( F^{(2)}_{\n\r]\a}- 
 F^{(1)\d}_{\n\r]} b_{\a\d} \right) - 2 a_\a H^{(1)}_{\m\n\r}\br 
&&\hskip 1.5in +2 m  B^{(1)}_{[\m\n}A^{(2)}_{\r]\a} )^2\br 
  && \hskip .5in  +6 ( F^{(3)}_{\m\n\a\b} + F^{(1)\d}_{\m\n} c_{\d\a\b}
+ 2 A_{[\m}^{(4)}\pa_{\n]} b_{\a\b} +2
\left\{ a_\a ( F^{(2)}_{\m\n\b}-b_{\b\d}F^{(1)\d}_{\m\n})-\left(
\a\leftrightarrow \b\right)\right\}\br
 &&\hskip1.5in  +  2 m \left\{ B^{(1)}_{\m\n} b_{\a\b} + A^{(2)}_{\m[\a} 
A^{(2)}_{\b]\n} \right\} )^2 \br
  && \hskip .5in  +4 (\pa_\m c_{\a\b\g} -2 \left(a_\a\pa_\m b_{\b\g} + 
{\rm cyclic\ perms.\ of \ \a,\b,\g}\right)\br &&\hskip 1.5in+ 2 m\ 
A^{(2)}_{\m[\a} b_{\b\g]})^2  + (2 m \ b_{[\a\b}b_{\g\d]})^2 ] + 
{1\over2} m^2 {\Large\}} \bigg], 
\label{17}\eea
where $\f={\h \F} -{1\over2} \ln G $ is  shifted dilaton field and the
scalars coming from $G_{\a\b}$ and $b_{\a\b}$ have been combined to form
the symmetric $8 \times 8$ matrix
\be
M= \eta M^{-1} \eta = \left( \begin{array}{cc} G^{-1} & -G^{-1} b \\ 
bG^{-1} & G-bG^{-1}b\end{array}\right), \ \ \ \  \ \eta=
\left(\begin{array}{cc} 0 & I_4 \\ I_4 & 0
\end{array}\right)
\la{18}\ee
where $\eta$ is $O(4,4)$ metric and $I_4$ is 4-dimensional identity matrix. 
We mention in passing that, if we had chosen to consider compactification on 
the d-dimensional torus, $T^d$,  then the corresponding $2d \times 2d$
symmetric M-matrix will appear, defined in terms of scalars coming from
the NS-NS sector and the metric $\eta$ for 
 $O(d,d)$ group with off diagonal identity matrix $I_d$ has to be introduced. 
Let us recall how various fields appear in the six dimensional action \eqn{17},
after the compactification. 
In the NS-NS sector we have
dilaton field, $\phi$, graviton, $g_{\m\n}$, tensor field,
$B^{(1)}_{\m\n}$, eight vector fields, coming from ten dimensional metric and
two index antisymmetric tensor fields after compactification and  
 sixteen scalar fields, ${\h g}_{\a\b}$
and ${\h B}_{\a\b}$, appearing in matrix M  which parameterize the coset
$ O(4,4)\ov O(4)\times O(4)$. On the other hand in the R-R
sector there are eight scalars from ${\h A}_\a$ and ${\h
C}_{\a\b\g}$, seven vectors from ${\h A}_\m$ and ${\h
C}_{\m\a\b}$, four 2-rank potentials from  ${\h
C}_{\m\n\a}$ and one 3-rank potential ${
C}_{\m\n\l}$. 
Let us recapitulate the symmetry of the six dimensional effective action for
the case when $m=0$ following the works of Sen and Vafa [26]. It was
shown in ref.26 that the action is invariant under $SO(4,4)$ symmetry
after the transformation properties of scalar, vector, and tensor fields
were defined. In fact the equations of motion are invariant under a larger
noncompact symmetry group $SO(5,5)$.  
On this occasion, the  massless case, one can combine dual of 
3-rank tensor field $C_{\m\n\l}$ with seven other RR vector fields to
form 8-dimensional spinorial representation, 
$\psi^a_\m (1\le a \le 8)$, of $SO(4,4)$. Similarly, 
3-form field strengths $G_{\m\n\l\a}$ can be taken to be
(anti)self-dual to form another 8-dimensional spinorial representation, 
$\psi^a_{\m\n\l}$. Note that eight RR-scalars do also transform under one 
of these spinor representation. The afore mentioned symmetry of massless 
six-dimensional theory was  exploited in [26] to generate type II 
dual pairs. However, one can explicitly check that in the case of dimensionally
reduced massive theory, the action is no longer invariant under the above
noncompact symmetry group. 

The  above action \eqn{17}  
is  invariant  under the following set of St\"uckelberg 
transformations (although it is tedius calculation),
\bea
&& \d A^{(4)}_\m= -m \ \L_\m, \br 
 && \d a_\a = -m \ \l_\a, \br 
&& \d b_{\a\b}=0, \br
&&\d A^{(1)\d}_\m=0, \ \ \
\d A^{(2)}_{\m\a}= \pa_\m \l_\a,\br
&&\d B^{(1)}_{\m\n}= \pa_{[\m}\L_{\n]} + F^{(1)\d}_{\m\n}\l_\d,\br
&&\d c_{\a\b\g}= -2m \ \l_{[\a} b_{\b\g]},\br
&&\d A^{(3)}_{\m\a\b}= -2m \ ( \L_\m b_{\a\b} +
A^{(2)}_{\m[\a}\l_{\b]}),\br 
&& \d B^{(2)}_{\m\n\a}= -2m \ (\L_{[\m} A^{(2)}_{\n]\a} + \l_\a
B^{(1)}_{\m\n}),\br
&&\d C_{\m\n\r}= -2m \ \L_{[\m} B^{(1)}_{\n\r]}.
\la{19}\eea

Here, $\Lambda_{\mu}$ and $\lambda_{\alpha}$ are vector and scalar gauge
functions respectively.
Note that in \eqn{19} RR-scalars do also transform under St\"uckelberg
transformations in lower dimensions. 

Next, we present black hole solutions in five and four dimensions. While 
looking for black hole solutions, we keep only dilaton and the two form field
strengths in the action \eqn{17}  and set the other scalar and tensor 
fields to zero. First we consider, the following D-dimensional action (in 
Einstein frame)
\bea
S_m= \int d^Dx \sqrt{-g}&& \bigg[ ( R_g - {4\ov D-2} \ \pa_\mu\f 
\pa^\mu\f -{1\over 2\cdot 2!}e^{{-4\ov D-2}\f}F_{\m\n}F^{\m\n} -2 
\l e^{{4\ov D-2}\f}) \br
&&-{1\over 2\cdot 2!}e^{{2(D-4)\ov D-2}\f}F_{R \m\n}F_R^{\m\n} -{1\over2} 
m^2 e^{{2D\ov D-2}\f} ], 
\la{ta}\eea
 We have added the term,  $\lambda e^{{4\over {(D-2)}}\phi}$, to the 
action and the presence of this term  can be interpreted as a dilatonic
potential which owes its origin from the NS-NS sector and
might appear due to some nonperturbative effects. The $m^2$ piece  comes
from the massive ten dimensional action (1) after compactification. In eq.(15)
the gauge field strengths  $F_{\mu\nu}$ and $F_{R \mu\nu}$  come from the NS-NS
and RR sectors respectively. 

The equations of motion  are
\bea
&&  {\nab_\m \nab^\m \f}  +{1\ov 8} e^{-{4\ov 
D-2}\f} F^2 -
{D-4 \ov 16} e^{{2(D-4)\ov D-2}\f}  F_R^2 
-{\l } e^{{4\ov D-2}\f} -{m^2 D \ov 8} e^{{2D\ov D-2}\f}=0, \br
&& ( R_{\m\n} -{1\ov2}g_{\m\n} R) - {4\ov D-2} (\pa_\m \f \pa_\n\f -{1\ov2} 
g_{\m\n}(\pa\f)^2) -
{1\ov 4} e^{-{4\ov D-2}\f} ( {2 F_{\m\l}F_\n^\l 
-{1\ov 2} g_{\m\n}F^2}) \br
&&-{1\ov 4} e^{{2(D-4)\ov D-2}\f} (2 F_{R \m\l}F_{R\n}^\l 
-{1\ov 2} g_{\m\n}F_R^2) + {1\ov 2} g_{\m\n}(2 \l e^{{4\ov D-2}\f} 
+{1\over2} m^2 e^{{2D\ov D-2}\f} ) =0 ,\br
&& \pa_\m e^{-4\ov D-2} F^{\m\n}=0,\br
&& \pa_\m e^{{2(D-4)\ov D-2}\f} F_R^{\m\n}=0.
\eea

We seek  maximally symmetric black holes   
solutions [28] and we choose constant 
dilaton backgrounds $\f=\f_c$ with the  metric ansatz 
\be
ds^2= - f(r) dt^2 +  f(r)^{-1} dr^2 + r^2 d\O_{{D-2}}^2,
\label{sol}
\ee
where $\epsilon_{D-2}$ and $d\O_{D-2}^2$ are the 
volume element and metric on  unit $S^{D-2}$, respectively.  
Our first example is a black hole with the following choice of 
backgrounds:  $F=0,~ F_R=0, with ~ \l<0$;  the solution \eqn{sol} is 
Schwarzschild-Anti-deSitter(SAdS) space with 
\bea &&e^{2\f_c}= {8  \ov 5}~{|\l|\ov m^2} 
\br && f(r)=  1 -{2 M\ov r^2} + {|\l|\ov 10} \left[ 
{8|\l|\ov 5 m^2}\right]^{2\ov 3} r^2
\eea
Note that  the black hole solution is asymptotically an AdS space with 
effective cosmological constant 
 $\L= {2 |\l| \ov 5}\left[ {8|\l|\ov 5 
m^2}\right]^{2\ov 3}$. 

The second example corresponds to the backgrounds: $F\ne 0,~F_R\ne 0 ~and 
~\l< 0$ , with the constraint  
\be 
Q_R^2 { m^2 \ov 2} = {8\ov 5} Q^2 |\l|,
\la{con}\ee
is satisfied  and the  charges are defined as 
\be
 Q= {1 \ov 2\pi^2} \int_{S^3} \ast e^{-{4 \ov 3} \f} F,\hskip .5in 
Q_R= {1 \ov 2\pi^2} \int_{S^3} \ast e^{ 
{2\ov 3}\f} F_R. 
\ee
The solution is 
the  Reissner-Nordstrom-AntideSitter(R-N-AdS) black hole in 5-dimensions 
\bea
&& e^{2 \f_h}= {1 \ov 2} {Q_R^2\ov Q^2}\br
&&
f(r)=  [ 1- ({r_+ \ov r})^2] [1- ({r_- \ov r})^2]  + {|\l|\ov 10} \left[ 
{8|\l|\ov 5 m^2}\right]^{2\ov 3} r^2
\br
&& \ast e^{-{4 \ov 3} \f} F= Q \epsilon_3,\hskip .5in \ast e^{ 
{2 \ov 3}\f} F_R= Q_R \epsilon_3,
\br && r_{\pm}^2= M \pm ( M^2 - {e^2 \ov 2} )^{1\ov 2},
\la{rnds}\eea 
 where M is a parameter, analog of mass (notice that
the space is not asymptotically flat)
and $e = {1\ov2} \bigg[ {Q_R^2\ov 
2} Q^2\bigg]^{1\ov 3}$ is related to the product of the two charges $Q_R$ and
$Q$ defined through eq.(20). 
It follows  from eqs. \eqn{con} and 
 \eqn{rnds} the string coupling at the black hole horizon is given by the
ratio of the two charges $Q_R$ and $Q$, and thus can be adjusted to be small
through the judicious choice of the ratio of the two charges. Note that  
 the spacetime in \eqn{rnds} is not 
asymptotically flat but has the curvature  equal to $5 \L$.
 We see from eq.(21) that near extremal blackhole solution can 
be obtained in the limit when $\lambda$ goes to zero and the two horizons 
come very close to each other, $i.e.$, $r_+ \sim r_-$.
Moreover, for $\l=0, m=0 $  above solution in \eqn{rnds} reduces to the 
Strominger-Vafa's five-dimensional extremal black hole solution [27]   
as expected. 

 We find a black hole solution in four dimensions for the  case when  only 
the 2-form RR-field strength is nonzero and as is well known for $D=4$, the
gauge field couples to gravity conformally. 
 The  black hole solution of the type \eqn{sol}  in four 
dimensions exists
  when $ F=0,~ F_R \ne 0,~\l<0 $. The solution is R-N-AdS with
\bea 
&& e^{2 \f_h}= {2 |\l|\ov m^2},\br
&& f(r) = ( 1-{ 2M \ov r} + {Q_R^2\ov r^2}) + \left[{\l^2\ov 
3m^2}\right]r^2,\br && \ast F_R = Q_R \epsilon_2,
\eea
where $Q_R= {1\ov 4 \pi}\int_{S^2} * F_R$.

Next  we turn our attentions to  obtain 4-brane solutions 
in six-dimensional model with
cosmological constant term in the RR sector, i.e. we set $\lambda = 0$ in 
this case. 
We choose the background field configurations in \eqn{17} so that
 except dilaton and
moduli matrix $M$ are nonvanishing and the resulting action takes the following form, 
\bea
S_m=\int d^6 x \sqrt{-g}&& \bigg[ e^{-2\f}\left(R + 4\pa_\m\f \pa^\m\f
+{1\over 8} Tr\pa_\m M^{-1}\pa^\m M \right)\br
&& -\sqrt{G}\left( {1\over 2\cdot 2!} ( m \ b_{\a\b} )^2  
+{1 \over 2 \cdot 3!} ( m b_{[\a\b}b_{\g\d]})^2
+ {1\over2} m^2 \right) \bigg],
\label{110}\eea
Note  that  with the introduction of mass term, $m$, in the ten 
dimensional effective action (1), the reduced action, with the specific 
choice of the  backgrounds, gets a piece which amounts to introducing a potential term 
involving the moduli $G_{\alpha\beta}$ and $b_{\alpha\beta}$, $\alpha ,\beta =
6,7,8,9$.
We seek for a four-brane solution around $b_{\alpha\beta}=0$
and, in the Einstein frame, 
$( g^E_{\m\n}= e^{-\f} g_{\m\n})$ the 4-brane solution is 
\bea
&& ds^2_E= U^{1\ov4} [ -dt^2 + dx_1^2 +dx_2^2+dx_3^2+dx_4^2]\ +
\ U^{5\ov4} dy^2,\br
&&e^{-{2\ov3} \f}= U^{1\ov2},\ \ \ \ U=\pm m |y-y_0|,\br
&& G_{\a\b}= \d_{\a\b}\ e^{{2\ov3}\f}, \ \ \ b_{\a\b}=0,
\la{111}\eea
 these  background configurations  satisfy 
all the equations of motion derived from \eqn{110}. 
This is a domain wall solution with a kink singularity (delta-function) at
$y=y_0$.
 The solution is not asymptotically flat, however,
for the choice, $ U =  \vert y - y_0 \vert$ at large distances, $e^{\phi}$
vanishes. This solution can be
oxidised to obtain 8-brane solution in ten dimensions. 

 To summarize our results: We considered ten dimensional effective action of 
type IIA theory in the presence of cosmological constant term which arises as
the dual of the ten dimensional field strength coming from the RR sector.
The action is dimensionally reduced on a d-dimensional torus with the
assumption that the fields do not depend on internal coordinates. The gauge 
invariance properties of the reduced action is investigated and the 
transformation properties of the fields in the NS-NS and RR sectors
are derived in the presence of the cosmological constant term. One of the 
interesting result is noticed in the six dimensional theory. It is found, that
in the case of the massive theory, in the presence of this cosmological
constant term,  the $SO(4,4)$ invariance is lost; whereas the massless theory
respects this symmetry.  
Thus, the cosmological constant coming from the RR
sector, in this case, breaks the T-duality symmetry: $SO(4,4)$. Moreover, it is
quite evident that, 
for the six dimensional massive theory, the equations of motion do not respect
the $SO(5,5)$ symmetry unlike the masless case [26]. 
We recall that
when one considers a four dimensional heterotic string theory and 
introduces a cosmological constant (in this case assumed to come from NS-NS
sector as central charge deficeit), the equations of motion do not respect
the S-duality invariance, as was discussed in ref.16. 
We presented, in this note, classical solutions of the effective action. In 
five dimensions we find black hole solutions in the presence of cosmological
constants. It is possible to get near extremal solutions for the choice of
small values of cosmological constant parameter, $\lambda$. In this 
context, we 
would like to point out that  our black holes are anti-de Sitter type and 
these solutions do not correspond to asymptotically flat 
case. Therefore, one has to define the Hawking temperature with some care. 
There have been attempts to understand thermodynamic properties of black holes
with (negative) cosmological constant term [29]. Brown, Creighton and 
Mann [30]
identify the thermodynamic internal energy of such a black hole and equate the
entropy to $ 1\over 4$ of the area of the black hole event horizon. The 
temperature on the boundary can be defined through thermodynamic relation 
between these two quantities, such that the black hole temperature, $T_H$,  is  
 $ {\kappa _H} \over{2\pi}$ times  the redshift
factor [31] for temperature in stationary gravitational field. The desired
relation is 
\bea 2\pi T(R) = {{\kappa _H}\over{{\cal N}(R)}} \eea
\noindent where ${\kappa _H}$ is the surface gravity at the 
horizon of the black hole and
${\cal N}(R) = \sqrt {-g_{tt}}$,  is the  lapse function. The temperature,
accordingly, depends on the location of the boundary. 
We have mentioned earlier that a massive type IIB effective action can be
obtained in nine dimensions from the ten dimensional type IIB theory through
generalised dimensional reduction due to Scherk and Schwarz. One can adopt
the toroidal compactification for that nine dimensional theory to obtain
reduced effective action in a way similar to the one presented recently [32]
and  explore various  symmetries 
in the massive theory. 

 We conclude this note with some speculations about the cosmological constant
problem and how the string symmetries might resolve it. We recall that for
the four dimensional string effective action, the equations of motion are not
invariant under S-duality when the cosmological constant is nonzero. In the
present case, we find that starting from the ten dimensional type IIA theory,
with the cosmological constant, when we consider the six dimensional theory
after dimensional reduction, the $SO(4,4)$ symmetry is broken. If we turn the
argument around, the $SO(4,4)$, a T-duality symmetry, if required to be a good
symmetry, will force us to set the cosmological constant to zero. Of course,
we are talking of toroidal compactification of type IIA to six dimension here
and in the case of four dimensional theory it was the heterotic string
effective action [16]. Nevertheless, it is quite amusing that in the two 
different cases, the constant is required to vanish (the symmetry 
requirements are different too). Therefore, it is quite tempting to 
conjecture
that the web of string dualities will impose strong constraints on the four
dimensional theory to tell us why the cosmological constant is vanishingly
small.
 
\noindent Acknowledgements: One of us (J.M.) would like to thank the members
of the Centre Physique Theorique,  for the warm 
hospitality where a part of this work was done.


\begin{thebibliography}{99}


\bibitem{1} E. Witten, `Some comments on String Theory Dynamics'; Proc. 
String '95, USC, March 1995, hep-th/9507121; Nucl. Phys. B443(1995)85,
hep-th/9503124.
\bibitem{2} P. K. Townsend, Phys. Lett. B350(1995)184;\\
 C. M. Hull and P. K. Townsend,
Nucl. Phys. B438(1995)109, hep-th/9410167.

\bibitem{3} For recent reviews see J. Polchinski, S. Chaudhuri and C. Johnson,
Notes on D-branes, Lectures at ITP, hep-th/9602052;\\
J. H. Schwarz, Lectures on Superstring and M-theory Dualities,
ICTP and TASI Lectures hep-th/9607201;\\
J. Polchinski, Lectures on D-branes,
in TASI 1996, hep-th/9611050;\\
 M. R. Douglas, Superstring Dualities and the Small
Scale Structure of Space, Les Houches Lectures 1996, hep-th/9610041;\\
P. K. Townsend, Four Lectures on M-theory, Trieste Summer School, 1996,
hep-th/9612121;\\
C. Bachas, (Half) A Lecture on D-brane, hep-th/9701019; \\
C. Vafa, Lectures on Strings and Dualities, hep-th/9702201.
\bibitem{4} J. H. Schwarz, Phys. Lett. B360(1995)13, hep-th/9510086; M-theory
Extensions of T-duality, he-th/9601077.
\bibitem{5} A. Sen, Int. J. Mod. Phys. A9(1994)3707;\\
A. Giveon, M. Porrati and E. Rabinovici, Phys. Rep. C244(1994)77;\\
E. Alvarez, L. Alvarez-Gaume and Y. Lozano, An Introduction to T-duality
in String Theory, Nucl. Phys. (proc. Supp.) 41 (1995) 1, hep-th/9410237. 
These review articles
cover various aspects of developments in the early phase of dualities.
\bibitem{6} A. Sen, Unification of String Dualities, hep-th/9609176.
\bibitem{7} J. Polchinski, Phys. Rev. Lett.75(1995)4724.
\bibitem{8} J. Polchinski and Y. Cai, Nucl. Phys. B296(1988)91;\\
C.G. Callan, C. Lovelace, C.R. Nappi and S.A. Yost, Nucl. Phys. 
B308(1988)221.
\bibitem{9} M.J. Duff and P. Van Nieuwenhuizen, \plb{94}{1980}{179},\\
A. Aurilia, H. Nicolai and P.K. Townsend, \npb{176}{1980}{509}.
\bibitem{10} L.J. Romans, Phys. Lett. B169(1986)374.
\bibitem{11} J.L. Carr, S.J. Gates, Jr and R.N. Oerter, Phys. Lett.
B189(1987)68.
\bibitem{12}J. Polchinski and E. Witten, {\it Evidence for heterotic
     type I duality}, \npb{460}{1996}{525}, \hepth{9510169}.
\bibitem{13} J. Polchinski and A. Strominger, {\it New vacua for
type II string theory}, \hepth{9510227}.
\bibitem{14} E. Bergshoeff, M. De Roo, M. Green, G. 
Papadopoulos and P. Townsend
, \npb{470}{1996}{113},\hepth{9601150}; E. Bergshoeff and M. B.
Green, {\it The type IIA super-eight brane}, preprint VG-12/95.
\bibitem{15} M.B. Green, C.M. Hull and P.K. Townsend, hep-th/9604119.
\bibitem{16} S. Kar, J. Maharana and H. Singh, Phys. Lett. B374(1996)43.
\bibitem{17} G. 't Hooft, Under the Spell of Gauge Principle, World Scientific
Publishing Co., Singapore, 1994, pa 352.
\bibitem{18} E. Witten, Mod. Phys. Lett. A10(1995)2153; Int. J. Mod. Phys. 
A10(1995)1247, \hepth{9506101}; Also see K. Becker, M. Becker and A. 
Strominger, Phys. Rev. D51 (1995) 6603, \hepth{9502107}.

\bibitem{19} J. Dai, R. G. Leigh and J. Polchinski, Mod. Phys. Lett.
A4(1989)2073;\\ M. Dine, P. Huet and N. Seiberg, Nucl. Phys. B322(1989)301.
\bibitem{20} J. Scherk and J. H. Schwarz, Nucl. Phys. B153(1979)61; Phys.
Lett. B82(1979)60.
\bibitem{21} E. Bergshoeff, C. M. Hull and T. Ortin, Nucl. Phys. B451(1995)547.
\bibitem{22} L. Andrianopoli, R. D' Auria, S. Ferrara. P. Fre, R. 
Minasian and M. Trigiante, hep-th/9612202.
\bibitem{23} P. Cowdall, H. Lu, C. N. Pope, K.S. Stelle and P. K. Townsend,
Nucl. Phys. B486(1997)49.
\bibitem{24} I. V. Lavrinneko, H. Lu and C. N. Pope, From topology to
generalised dimensional reduction, CTP-TAMU-59/96; hep-th/9611134.
\bibitem{25} J. Maharana and J. H. Schwarz, Nucl. Phys. B390(1993)3. For
similar works on string compactifications see: F. Hassan and A. Sen, Nucl.
Phys. B375(1992)103; S. Ferrara, C. Kounnas and M. Porrati, Phys. Lett.
B181(1986)263; M. Terentev, Sov. J. Nucl. Phys. 49(1989)713.
\bibitem{26} A. Sen and C. Vafa, 
               \npb{455}{1995}{165}.
\bibitem{27} A. Strominger and C. Vafa, \plb{379}{1996}{99} \hepth{961029}.
\bibitem{28} G. Gibbons and K. Maeda, \npb{298}{1988}{741}.
\bibitem{29} S. W. Hawking and D. N. Page, Commun. Math. Phys. 87 (1983) 577.
\bibitem{30} J. D. Brown, J. Creighton and R. B. Mann, Phys. Rev. 
D50 (1994) 6394; J. D. Brown and J. York, The path integral formulation 
of gravitational thermodynamics, IFP-UNC-491, CTMP/007/NCSU.
\bibitem{31} R. C. Tolman, Phys. Rev. 35(1930)904.

\bibitem{32} J. Maharana, S-duality and Compactification of type IIB 
Superstring action, Phys. Lett. B(in press), hep-th/9703009; Shibaji Roy, 
On S-duality of toroidally compactified type IIB string effective action,
 preprint SINP-TNP/97-02,
\hepth 9705016.

\end{thebibliography}
\end{document}